\documentclass[11pt]{article}
\usepackage{amsmath}
\usepackage{graphicx}
\usepackage{amsfonts}
\usepackage{amssymb}
\usepackage{epsf}
\usepackage{latexsym}

\def\80{\hspace{0.8in}}

\newcommand{\be}{\begin{enumerate}}
\newcommand{\ee}{\end{enumerate}}
\newcommand{\bi}{\begin{itemize}}
\newcommand{\ei}{\end{itemize}}
\newcommand{\bd}{\begin{description}}
\newcommand{\ed}{\end{description}}

\def\beq{\begin{equation}}
\def\eeq{\end{equation}}
\def\bea{\begin{eqnarray}}
\def\eea{\end{eqnarray}}
\def\pa{\partial}
\def\d{\textrm{d}}

\def\mL{\mbox{L}}

\def\mN{\mbox{N}}

\def\sd{\mbox{\scriptsize d}}

\def\sB{\mbox{\scriptsize B}}

\def\sK{\mbox{\scriptsize K}}

\def\sN{\mbox{\scriptsize N}}

\def\sW{\mbox{\scriptsize W}}

\def\eph(B){\mbox{\scriptsize emergent(LMB)}}

\def\tN{\mbox{\tiny N}}

\def\fE{\mbox{\sffamily E}}

\def\fT{\mbox{\sffamily T}}

\begin{document}
\begin{titlepage}
\vspace{.7in}
\begin{center}
 
\vspace{2in} 

\LARGE{\bf SMALLEST RELATIONAL MECHANICS MODEL}

\vspace{.1in}\LARGE{\bf OF QUANTUM COSMOLOGY}

\vspace{.2in}

\large{Edward Anderson$^{1}$} 

\vspace{.2in}

\large{\em $^1$ DAMTP Cambridge U.K.}

\vspace{.2in}

\end{center}

\begin{abstract}

Relational particle mechanics are models in which there is, overall, no time, position, orientation 
(nor, sometimes, scale).  
They are useful for whole-universe modelling - the setting for quantum cosmology.  
This note concerns 3 particles in 1d in shape-scale split variables.  
The scale part parallels certain Friedmann equations, while in this note the shape part involves 
functions on the circle.  
The scale part is taken to be `heavy' and `slow' so the semiclassical approach applies and scale 
provides an approximate timestandard with repect to which the light physics runs.  
Relational particle mechanics moreover provide conceptual models of inhomogeneity, structure formation 
and nontrivial linear constraints (minisuperspace models do not and midisuperspace models only 
do at the cost of substantial complications).

\end{abstract}

\vspace{0.2in}


\vspace{3.2in}

\noindent$^1$ ea212@cam.ac.uk \mbox{ }

\end{titlepage}

\section{Introduction}

Euclidean relational particle mechanics (RPM) \cite{BB82, Cones} has no absolute time, absolute position 
or absolute orientation (in the rotational sense).  
Similarity RPM is likewise \cite{B03, FORD, AF, 08I} but also with no absolute scale.  
RPM's are useful toy models of GR in a number of ways (listed in \cite{AF}) resembling it 
(particularly in the formulations \cite{RWRABFKO}) to a comparable but different extent to the more 
habitually studied minisuperspace models. 
In addition to having a constraint quadratic in the momenta that leads to the frozen quantum equation 
aspect of the Problem of Time \cite{POT}, RPM's (unlike minisuperspace) have nontrivial constraints 
that are linear in the momenta.
In the case of Eulidean RPM, the latter is a zero total angular momentum constraint whereby one passes 
to the quotient of relative particle separations by rotations.  
These parallel the GR momentum constraint, its association with spatial diffeomorphisms and the quotient 
superspace of the space of Riemannian 3-metrics by these.   
Also RPM's (unlike minisuperspace) have notions of locality and clustering, making them more amenable to 
records-theoretic modelling \cite{Records} and some aspects of semiclassical quantum cosmological 
modelling \cite{HallHaw, POT}. 

\mbox{ }

This note addresses the semiclassical quantum cosmology aspect.    
It concerns a regime in which there is a split into heavy, slow variables and light fast variables (in 
the Born-Oppenheimer and WKB senses respectively).  
In GR, these are, respectively, homogeneous and inhomogeneous modes. 
In Euclidean RPM, these are scale (moment of inertia of the model universe) and shape (clustering).  
The heavy slow variables provide an (approximate) timestandard with respect to which the light fast 
variables evolve.  

\mbox{ }

RPM's are straightforward to study in 1 or 2 spatial dimensions: for $N$-particle similarity RPM, these 
have configuration spaces $\mathbb{S}^{N - 2}$ and $\mathbb{CP}^{N - 2}$ respectivly \cite{FORD}.  
For Euclidean RPM, one has the corresponding cones \cite{Cones}.  
In 1-d with N particles, scale comes in via radius $\iota =\sqrt{I}$ (for $I$ the moment of inertia of 
the system), so one has a Cartesian interpretation (the 2-d 3-particle sphere $\mathbb{CP}^1 = 
\mathbb{S}^2$ is harder in these respects \cite{Cones}).   
This note covers the 3-particle case.

\section{The smallest RPM model of quantum cosmology}

In terms of 2 mass weighted Jacobi coordinates $\iota_1$, $\iota_2$ and then passing to the 
corresponding polar coordinates $\iota$, $\Phi$, the relational kinetic term for 3 particles on a line 
becomes $\fT = \{\dot{\iota^2} + \iota^2\dot{\Phi}^2\}/2$, while this note's multi-HO potential maps to 
$\iota^2\{A + B\mbox{cos}2\Phi\}$ for $A = \{K_1 + K_2\}/4> 0$ and $B = \{K_1 - K_2\}/4$ of whatever 
sign (for $K_1$, $K_2$ the Jacobi--Hooke coefficients divided by the Jacobi cluster masses). 
We only consider the case with a spring between particles 1 and 2 and an effective spring between 
particle 3 and the centre of mass of particles 1 and 2 (else there would be a third potential term).  

\mbox{ }

The scale part of this obeys the analogue Friedmann equation 
\beq
\left(\frac{\dot{a}}{a}\right)^2 = \frac{2\fE}{a^2} - 2A
\label{Fried}
\eeq
i.e. with a `curvature' term and a `cosmological' constant term, and so corresponding to the Milne in 
AdS model \cite{Rindler}.  

\mbox{ }

Then 1) the $B = 0$ case at the QM level with conformal ordering \cite{Banal} is mathematically a 2-d 
isotropic HO, which has a standard solution, which in terms of our problem's original quantities, takes 
the following guise.  
The eigenvalues are 
\beq
\frac{\sqrt{2}\fE}{\hbar\sqrt{K_1 + K_2}} = |\d| + 2\mN + 1 \mbox{ } , 
\eeq
for d a relative dilation quantum number and N a quantum number that counts the nodes in the  
`radius' $\iota$, and the wavefunctions are   
\beq
\Psi_{\tN\sd}(\iota, \Phi) \propto \iota^{|\sd|}\mbox{exp}
\left(
-\frac{\sqrt{K_1 + K_2}\iota^2}{2\sqrt{2}\hbar}
\right)
\mL_{\sN}^{|\sd|}
\left(
\frac{\sqrt{K_1 + K_2}\iota^2}{\sqrt{2}\hbar}
\right)
\mbox{exp}(i\d \Phi) \mbox{ } .
\eeq
2) Even $B \neq 0$ is exactly soluble by using Cartesian coordinates, in terms of which one obtains 
the usual Gaussian times Hermite polynomial form in each Cartesian coordinate.  
One can then dress this answer up in our problem's significant shape-scale variables.  

\mbox{ }

\noindent 3) In parallel to the situation in cosmology, we declare $\iota = H$ and $\Phi = L$.
Then the H-equation [which amounts to (\ref{Fried})] gives the approximate emergent WKB timestandard 
(= cosmic time here) 
\beq
t^{\sW\sK\sB} =   
\sqrt{\frac{2}{K_1 + K_2}}\mbox{arcsin}\left(\sqrt{\frac{\{K_1 + K_2\}}{4\fE}}\iota\right)  \mbox{ } .  
\eeq
Next, the L-equation is, in terms of the rectified time 
$$
T = - \sqrt{\frac{2}{K_1 + K_2}}\mbox{cot}
\left(
\sqrt{    \frac{K_1 + K_2}{2}    }t^{\sW\sK\sB}  
\right) \mbox{ } ,
$$ 
\beq
i\hbar\frac{\pa|\chi\rangle}{\pa T} = - 
\frac{\hbar^2}{2M}\frac{\pa^2|\chi\rangle}{\pa\Phi^2} + 
\frac{  BM\,\mbox{cos}\,2\Phi  }{  \{1 + \{K_1 + K_2\}T^2/2\}^2  }  |\chi\rangle 
\mbox{ } ,
\label{main}
\eeq
for `mass' $M = 4\fE/\{K_1 + K_2\}$, which, for $B$ small, i.e. $K_1 \approx K_2$, poses, about a very 
simple QM equation, a (fairly analytically tractable) $T$-dependent perturbation problem.  
This parallels Halliwell--Hawking's work \cite{HallHaw} while being simpler and so permitting rather 
more and rather more straightforward checks of various Problem of Time and quantum cosmological ideas.

\section{Conclusion}

\noindent 
This note's analysis extends in many ways to the $N$-particle case and to tighter analogies with more 
commonly encountered cosmological models \cite{Cones, SemiclIII}.  
The reason for studying this note's `negative curvature balanced by negative cosmological constant' 
type scenario is that it it is ulteriorly exactly soluble \cite{SemiclI}, permitting in this case what 
are usually unavailable checks of the circumstances under which the various assumptions and 
approximations used in the semiclassical approach to the problem of time and quantum cosmology do and do 
not hold for the exactly solved model. 
This is work in progress \cite{SemiclIII}.

\mbox{ }

\noindent
To incorporate nontrivial constraints, 2-d models are considerably more useful \cite{08I, 08III}; 
however, these models are harder, and so I am considering $N$ particles on the line first.  
Records-theoretic study is best left to models with 4 or more particles.   

\mbox{ }

\noindent{\bf Acknowledgments}

\mbox{ }

\noindent This material was presented at the Marcel Grossmann Meeting, Paris, 2009.  

\vspace{2in}


\end{document}